\documentclass[submission,copyright,creativecommons]{eptcs}
\providecommand{\event}{EPTCS 2023} % Name of the event you are submitting to

\usepackage{iftex}

\ifpdf
  \usepackage{underscore}         % Only needed if you use pdflatex.
  \usepackage[T1]{fontenc}        % Recommended with pdflatex
\else
  \usepackage{breakurl}           % Not needed if you use pdflatex only.
\fi

\usepackage{graphicx}
\usepackage{amsmath}

\title{Metatickles and Death in Damascus}
\author{Saira Khan
\institute{University of California, Irvine\\ Irvine, California}}

\def\titlerunning{Metatickles and Death in Damascus}
\def\authorrunning{S. Khan}

\begin{document}
\maketitle

\begin{abstract}
 The prescriptions of our two most prominent strands of decision theory, evidential and causal, differ in a general class of problems known as Newcomb problems. In these, evidential decision theory prescribes choosing a dominated act. Attempts have been made at reconciling the two theories by relying on additional requirements such as ratification (\cite{jeffrey1983}) or ``tickles'' (\cite{eells1982}). It has been argued that such attempts have failed (\cite{lewis1981a}; \cite{skyrms1982}). More recently, Huttegger (\cite{huttegger2023}) has developed a version of deliberative decision theory that reconciles the prescriptions of the evidentialist and causalist. In this paper, I extend this framework to problems characterised by decision instability, and show that it cannot deliver a resolute answer under a plausible specification of the tickle. I prove that there exists a robust method of determining whether the specification of the tickle matters for all two-state, two-act problems whose payoff tables exhibit some basic mathematical relationships. One upshot is that we have a principled way of knowing \textit{ex-ante} whether a reconciliation of evidential and causal decision theory is plausible for a wide range of decision problems under this framework. Another upshot is that the tickle approach needs further work to achieve full reconciliation.
\end{abstract}

\section{Introduction}

Decision theory offers a normative framework for determining rational choice. Its primary components are a set of beliefs (probabilities) over states of the world and a set of valuations (utilities) over the different outcomes of acts in these states of the world. Two prominent forms of decision theory are the causalist and the evidentialist approaches. Causal decision theory determines rational action by evaluating what an agent can expect to bring about by her action. Evidential decision theory determines rational action by evaluating what evidence an agent’s action provides her with.

The theories prescribe different acts as rational under a class of problems known as Newcomb problems. It is frequently held that the causalist prescription is the correct one (\cite{stalnaker1968}; \cite{gibbardharper1978}; \cite{skyrms1980}; \cite{lewis1981a}).\footnote{Though some, such as \cite{ahmed2014}, \cite{horgan1981} and \cite{horwich1985} support the evidentialist conclusion.} The characteristic feature of Newcomb problems is that there is a correlation between state and act such that the choosing of the act is understood to be good evidence for a state of the world. The result is that evidentialism prescribes choosing an act which is strictly worse in both states of the world. The evidentialist recognises that though the agent cannot causally bring about a different state of the world, they deny that causality is important for practical rationality (\cite{ahmed2014}). Rather, the rational act should be based on its ``news value''. That is, an agent ought to prefer a proposition to another just in case she would rather learn that proposition over the other. In light of criticism of this position, attempts have been made -- notably by Jeffrey (\cite{jeffrey1983}) and Eells (\cite{eells1982}) -- to amend evidential decision theory to better accord with causalist prescriptions. 

In this paper, I focus on a version of reconciliation developed by Huttegger (\cite{huttegger2023}) and show that it cannot reconcile evidential and causal decision theory without further, questionable assumptions. Huttegger uses an idea due to Eells called the ``tickle'' defence: that the evidentialist becomes increasingly confident that the state of the world is not causally dependent on her act as a result of knowledge of her beliefs and desires. However, Huttegger employs the deliberative apparatus developed by Skyrms (\cite{skyrms1982}) and thus overcomes some objections to the original Eellsian approach.\footnote{In particular, the assumption that the agent access to a proposition which fully describes her beliefs and desires. Under Huttegger's approach, this is not assumed but rather reached through a process of deliberation.} Section 2 of this paper expounds the technical differences between causal and evidential decision theory and briefly outlines two decision problems: the Newcomb problem and Death in Damascus. Section 3 discusses Eells' approach to resolving the difference between the evidentialist and causalist prescriptions and details Huttegger's proposed amendment using deliberative dynamics. Huttegger's approach delivers the (commonly considered) correct answer for the evidentialist in the Newcomb problem.

Section 4 considers the same framework applied to a class of problems characterised by decision instability. These are where, as soon as the agent leans toward performing one action, the other looks preferable. In more technical terms: there is no dominant act (no act which is preferred regardless of the state of the world) and every act is in principle causally unratifiable (after we have chosen the act we would prefer to have chosen otherwise). In particular, I consider a decision problem known as Death in Damascus (\cite{gibbardharper1978}). When the payoff table is symmetric, the received view is that both naïve evidentialism and naïve causalism (without any deliberative dynamics) remain silent on which is the correct act to perform. When it is asymmetric, the evidentialist is decisive whereas the causalist is trapped in a state of indecision. A more sophisticated (deliberative) causalist may settle upon choosing an act with probability slightly less than 0.5. In this paper, we see that Huttegger's framework, when applied to this problem, cannot straightforwardly reconcile the evidentialist prescription with the prescription of the causalist (both sophisticated and naïve).

In Section 5, I offer an original analysis of the deliberative framework to explicate why it is irresolute in the Death in Damascus problem, and prove some general facts about its irresoluteness given a plausible version of the dynamical process, which I call the \textit{shortest-path independence dynamics}. I identify the existence of what I call the \textit{plane of indifference} in all two-act, two-state decision problems which exhibit the basic mathematical structure of either Newcomb or Death in Damascus problems. The key insight is that the specification of the tickle matters only depending on the positioning of this plane of indifference. In particular, regardless of the precise operation of the tickle during deliberation -- shortest-path or not -- the positioning of the plane in the Newcomb problem renders it the case that deliberation will always lead us to the same conclusion. This is not so in Death in Damascus and reconciliation of evidential and causal decision theory here requires more questionable assumptions. Section 6 discusses the status of reconciliation and the importance of the proof of the indifference plane for future work in deliberative decision theory. Section 7 concludes an offers a view on the status of the Eellsian project.

\section{The decision problems}

The canonical form of evidential decision theory is attributable to Jeffrey (\cite{jeffrey1983}). Under his framework, states of the world, acts and outcomes are all propositions of the same kind, forming a Boolean algebra. Probabilities and desirabilities may be applied to any of these propositions. Call the Boolean closure of the set of acts, states and outcomes, the \textit{decision-relevant} propositions. The agent's conditional expected utility of an act is calculated from her probabilities and desirabilities for maximally specific decision-relevant propositions. Formally, the evidential decision theorist prescribes performing the act, $A$, that maximises the following conditional expected utility formula, where $D$ denotes desirability, $P$ denotes probability, and $S$, the state of the world.

$$EU_{evid}(A) = \displaystyle\sum_{i} D(S_i \& A)P(S_i|A)$$

There are multiple versions of causal decision theory.\footnote{Most notably, the subjunctive accounts of Stalnaker (\cite{stalnaker1968}) and Gibbard and Harper (\cite{gibbardharper1978}), as well as the non-subjunctive accounts of Skyrms (\cite{skyrms1980}) and Lewis (\cite{lewis1981a}).} For simplicity, I present Lewis' (\cite{lewis1981a}) account. Like the traditional decision-theoretic framework of Savage (\cite{savage1954}), states, acts and outcomes are not propositions of the same Boolean algebra but are separate entities. Probabilities attach to states of the world, and desirabilities or utilities, to outcomes. Lewis builds on the Savage framework but introduces \textit{dependency hypotheses} which determine the appropriate partition of the state space. A dependency hypothesis is defined as the maximally specific proposition about how outcomes do, and do not, causally depend on the agent's present acts. Formally, the causal decision theorist prescribes performing the act, $A$, that maximises the following expected utility formula relative to the partition given by the dependency hypothesis.\footnote{The merit of the evidential approach is that it is partition invariant and it is much less sensitive to the formal specification of the decision problem. Indeed, it is a more general framework that can be reduced to Savage's decision theory under correct specification of the state space. In comparison, in many causal decision theories, the decision problem must be specified in such a way that each state-act pair is guaranteed to lead to a unique outcome; there is state-act independence; and the desirabilities of the outcomes are not influenced by the state-act pair which eventuated them. None of these restrictions are required in the evidential framework. See Eells (\cite{eells1982}) for discussion.} 

$$EU_{caus}(A) = \displaystyle\sum_{i} D(S_i \& A)P(S_i)$$

I now present two decision problems. One which has caused particular worry for the evidentialist, and one which has caused worry for both theories, though it is more frequently levied against the causalist (\cite{egan2007}). The first, Newcomb's problem, can be described as follows (\cite{nozick1969}). Tomas is in a room with two boxes, one of which is opaque and one of which is transparent. Under the transparent box lies \$1,000. Under the opaque box, there is either nothing or \$1,000,000 and Tomas does not know which. He is offered the option to take either only the opaque box, or both the transparent one and the opaque one. The catch is that there is a predictor who, if she predicts Tomas chooses only the opaque box puts \$1,000,000 under it and, if she predicts he chooses both boxes, puts nothing under it. Tomas believes the predictor is reliable. The payoff table is illustrated Table 1.

\begin{table}[h]
    \centering
        \begin{tabular}{ l | l | l}
            & Box empty & Box not empty \\
            \hline
            Take opaque box & 0 & 1,000,000 \\
            \hline
            Take both boxes & 1,000 & 1,001,000 \\
        \end{tabular}
        \caption{Newcomb's Problem}
\end{table}

In this decision problem, the causalist recommends taking both boxes, as it can be seen that this act strictly dominates taking only the opaque box. That is, it has higher expected utility under both states of the world. The naïve evidentialist, however, recommends taking only the opaque box, as choosing only the opaque box is good evidence that the predictor put \$1,000,000 there. In this decision problem, the evidentialist seems to prescribe the wrong answer and Tomas loses out on a guaranteed \$1,000. 

The Death in Damascus problem is as follows (\cite{gibbardharper1978}). Death works from an appointment book which specifies a time and a place. If and only if Tereza happens to be in the time and place when Death is there, she dies. Suppose Tereza is in Damascus and she accidentally bumps into Death. He tells her that he is coming for her tomorrow. Her options are either to stay where she is or to flee to Aleppo. The catch is that Death is a reliable predictor of where she will be, so as soon as Tereza believes it is better for her to flee, this constitutes good evidence that Death's appointment for her is in Aleppo and it seems as though she should stay. Analogously, however, if she decides to stay, this constitutes good evidence that Death knows that she stays and so she would be better off fleeing. The problem is therefore one of decision instability. The moment Tereza becomes confident in one option, the other appears more attractive. Here, I consider an asymmetric problem where the cost of fleeing is 1 util. The payoff table is given in Table 2, where we assign 10 utils to Tereza's survival.\footnote{While only the asymmetric case is presented in this paper,  for completeness, the symmetric case was also analysed. This exhibits multiple lines of equilibria on the faces of the dynamical cube and therefore constitutes greater instability on the boundary than the asymmetric case. However, some would deny that indecision in such a circumstance constitutes a flaw in the theory. See, for example, \cite{harper1984}.}

\begin{table}[h]
    \centering
    \begin{tabular}{ l | l | l}
    & Death in Damascus & Death in Aleppo \\
    \hline
    Stay in Damascus & 0 & 10 \\
    \hline
    Flee to Aleppo & 9 & -1 \\
    \end{tabular}
    \caption{Asymmetric Death in Damascus Problem}
\end{table}

In this decision problem, the naïve evidentialist believes that, as Tereza's act is good evidence of the state of the world no matter what she chooses, she ought to stay in Damascus, since she should not pay the extra 1 util to flee to Aleppo. The causalist, however, believes that staying is irrational as it will put the agent in a position from which fleeing looks superior. She is therefore in a state of decision instability. Gibbard and Harper (\cite{gibbardharper1978}) argue that this is the correct answer as neither choice is ratifiable. Other forms of causal decision theory, for example, the deliberative framework of Joyce (\cite{joyce2012}) or Arntzenius (\cite{arntzenius2008}), prescribe the mixed act of fleeing with probability $0.474$.\footnote{This is derived using Joyce's (\cite{joyce2012}) framework for Murder Lesion applied to Death in Damascus assuming conditional probabilities $P(S2|A2) = P(S1|A1) = 0.99$. Under this framework, one's unconditional probabilities are revised in light of the expected utility calculation of an act in conjunction with the probabilistic correlation between state and act. More precisely, let $\alpha$ be a real number, $P_{t+1}(S2) = P_t(S2|EU_t(A2) = \alpha) \neq P_t(S2)$ when $\alpha \neq 0$, so the probability of a state of the world is updated based on its probability conditional upon the expected utility of an act. Further, let $x$ and $y$ be real numbers, if $P_t(A2) < 1$ and $x > y$, then $P_t(A2|EU_t(A2) = x \hspace{1mm} \& \hspace{1mm} EU_t(\sim A2) = y) > P_t(A2)$, so the choice probability of an act is updated based on its expected utility. The iterative process of updating one's choice probability will continue in this fashion until $P_t(A2) = P_{t+1}(A2) = P_t(A2|EU_t(A2))$, so information about its expected utility does not change its choice probability. As in Skyrms' (\cite{skyrms1982}) deliberational framework, this occurs when the expected utility of the two acts are equal.} In Skyrms' and Huttegger's deliberative dynamics, the agent only has access to pure acts and is therefore in a state of indecision when deliberation assigns an act probability of less than 1. In Joyce's framework, the mixed act is a choice for the agent should she have access to a random chance device she may use to pick her final, pure act. That is, a chance device which will determine that she flees with probability $0.474$. One might ask whether the evidentialist should be reconciled with the naïve causalist or deliberative causalist. If we sought similar instability as the naïve causalist, it will be clear from the analysis which follows that this will not be achieved: in many cases the deliberative evidentialist is decided. So I ask whether reconciliation with the Joycean causalist is possible on Huttegger's model -- whether evidential decision theory can prescribe the mixed act of fleeing with probability $0.474$. First, we must explicate the framework.

\section{A brief history of the metatickle approach and Huttegger's dynamics}

A prominent evidentialist attempt to prescribe the causalist action in the Newcomb problems is attributable to Eells (\cite{eells1982}; \cite{eells1984}). This has been referred to as the ``tickle'' or ``metatickle'' defence (\cite{lewis1981a}; \cite{skyrms1982}).\footnote{It is so named for the following thought experiment. Suppose the agent feels a tickle in his left pinkie just in case the predictor has put \$1,000,000 in the opaque box. Then, even though the presence of money depends probabilistically on the agent's act, the tickle is sufficient to screen off the relevance of that act to the state of the world -- the tickle tells the agent all he needs to know. A tickle may not always be available but, according to Eells, a ``metatickle'' is. This is a proposition which describes the agent's beliefs and desires.} Eells argues that the mistake being made by the naïve evidentialist in the Newcomb problem is the inference from some underlying common cause of both state and act, to a dependence \textit{of} the state \textit{on} the act. Eells argues that the only way in which the underlying cause could affect an agent's act is through the agent's beliefs and desires since, under our decision theories, these are the entities that determine action.\footnote{Eells suggests the common cause could not affect an agent’s act by changing his decision rule. In particular ``the agent believes that the causal influence of the common cause is sufficiently insignificant as to be irrelevant to the eventual determination of which act is correct in light of his beliefs and desires... This is because he believes that the causal influence of whatever is causally responsible for his rationality -- his training, genetic make-up, and so on – will be overwhelming'' (\cite[147]{eells1982}).} This implies that if the agent had full knowledge of his beliefs and desires, knowledge of the presence or absence of the common cause would be irrelevant to his act.

The intuition is clear with a simple example. Consider a decision problem with the same structure as the Newcomb problem but is instead a decision about whether or not to smoke cigarettes. Suppose that there is a genetic cause, $C$, that results in both lung cancer and a proclivity to enjoy cigarettes but smoking does not itself result in lung cancer. It is correlated with lung cancer but there is no causal state-act dependence. Causal decision theory recognises this independence and thus prescribes smoking insofar as it is enjoyable to the agent. The naïve evidentialist prescribes abstaining as smoking is good evidence for the presence of the gene which determines lung cancer. The Eellsian evidential decision theorist, however, believes that the only way the common cause can affect the agent's acts is through his beliefs and desires. Let the proposition which describes his beliefs and desires be denoted $T$ for metatickle. We have:

$$P(A | T \& C) = P(A | T \& \sim C)$$
If an agent has full knowledge of her beliefs and desires, $P(T) = 1$. So in the presence of the metatickle,
$$P(A | C) = P(A| \sim C)$$
By symmetry of probabilistic independence, 
$$P(C | A) = P(C | \sim A)$$
Since the cause is not probabilistically dependent on the act in the presence of the metatickle, neither is the state of the world. This means
$$P(S | A \& T) = P(S | T)$$

\begin{figure}[h]
    \centering
    \includegraphics[scale=0.65]{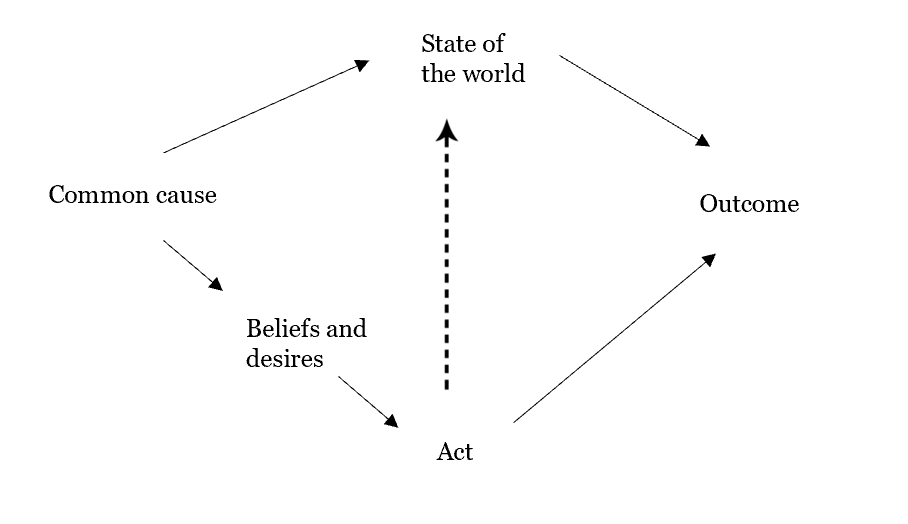}
    \caption{Diagrammatic depiction of Eells’ metatickle defence where the causal connection between act and state is erroneously drawn on the basis of the common cause}
    \label{fig:eells}
\end{figure}

Eells believed that the proposition $T$ was a proposition available to an agent (\cite{eells1982}; \cite{eells1984}). Conditional upon $T$, state and act are independent, and if this is the case, evidential decision theory will make the correct prescription: to smoke. Knowledge of the beliefs and desires of the kind caused by the common cause screens off what was previously thought to be evidence about the state of the world: the act. Analogous reasoning will lead the Eellsian evidential decision theorist to two-box in Newcomb's problem; the act of two-boxing is irrelevant to the \$1,000,000 being there or not, and one should therefore choose the strictly dominant act.\footnote{See also Reichenbach's principle of screening off (\cite{reichenbach1959}).} The reasoning behind the metatickle approach is diagrammed in Figure~\ref{fig:eells}.

For both Eells and Jeffrey (\cite{jeffrey1981}), it is the agent's ability to anticipate her own choices that screens off the evidential import of her acts for states of the world.\footnote{Indeed, Skyrms (\cite[74]{skyrms1984}) refers to Jeffrey's idea as a ``hypothetical version of the metatickle defense''.} However, unlike Eells, Jeffrey does not make reference to common causes. For Jeffrey, deliberation is what allows the sophisticated evidentialist to screen off the correlation between act and state which caused her to disagree with the causalist. He states ``it is my credences and desirabilities at the end of deliberation that correspond to the preferences in the light of which I act, i.e., it is my final credence and desirability functions [...] not the initial ones [...] that underlie my choice'' (\cite[486]{jeffrey1981}). The idea is that the agent should not choose to maximise news value as she now sees it, but as she now expects herself to estimate it after having made the decision. This is known as ``ratificationism''. However, Huttegger believes that both Eells and Jeffrey did not adequately specify how the agent comes to fully know her beliefs and desires and achieve this screening off (\cite{huttegger2023}).\footnote{See also \cite{lewis1981a}; \cite{ahmed2014}; \cite{horwich1985}; \cite{joyce1999}; \cite{skyrms1982}; \cite{skyrms1984} for criticisms of the metatickle approach.} To fill this lacuna, he first turns to the deliberational dynamics of Skyrms (\cite{skyrms1982}).

Skyrms models a deliberational process where, as an agent deliberates about which act to choose, this is incorporated into her up-to-date probabilities and desirabilities. The agent has some information at the start of deliberation upon which she can assess expected utility but the deliberation process itself generates information that causes her to recalculate her expected utility. Suppose we assign probabilities to acts that represent the agent’s belief that she will choose that particular act at the end of deliberation. Since states and acts are correlated, act probabilities provide evidence about states of the world which the agent can use to update her expected utility. Deliberation then pushes the agent in the direction of the act that has the higher expected utility in his current assessment. In particular, the direction of his choice probability of choosing both boxes, denoted $P(A2)$, is proportional to the difference in expected utility so that we have:

$$\frac{dP(A2)}{dt}  \propto EU(A2) - EU(A1)$$ 

And 

$$
\frac{dP(A2)}{dt} = 
\begin{cases}
\text{positive if} \hspace{1mm} EU(A2) > EU(A1)\\
\text{negative if} \hspace{1mm} EU(A2) < EU(A1)\\
\text{$0$ if} \hspace{1mm} EU(A2) = EU(A1)
\end{cases}
$$

We will refer to this as the ``adaptive dynamics''.\footnote{Skyrms also refers to this informally as a dynamical rule which ``seeks the good'' (\cite[30]{skyrms1990}). He describes such rules as ``qualitatively Bayesian'' in the sense that the dynamical rule should reflect the agent's knowledge that she is an expected utility maximiser and the status of her present expected utilities as an expectation of her final utilities. Informally, such rules state that act probabilities should increase if the act has utility greater than the status quo, and that the probability of all acts with utilities greater than the status quo should increase. Frequently used dynamical rules that meet these conditions are the replicator dynamics or Nash dynamics, and the dynamics of Brown and von Neumann (\cite{skyrms1990}). Formally, $\frac{dP(A)}{dt} = \frac{cov(A) - P(A)\sum_j cov(A)_j}{k + \sum_j cov(A)_j}$ and $\frac{dP(A)}{dt} = cov(A)^2 - P(A)\sum_j cov(A)_j^2$ respectively, where the constant $k$ represents how quickly the agent adjusts her act probabilities.}. It is assumed, in both Skyrms' and Huttegger's frameworks, that the adaptive dynamics operates continuously, though others, such as Eells \cite{eells1984} have developed discontinuous approaches. Since this paper is engaging with Huttegger's reconciliation project, I will assume a continuous adaptive dynamics. For Skyrms, the updating of one's choice probability continues until such a time as the agent reaches probability $1$ of performing a certain act or the agent reaches a mixed equilibrium where there is no change in her choice probabilities ($\frac{dP(A2)}{dt} = 0$). The basic intuition capturing the metatickle is that, if Tomas leans toward only taking the one box, the probability of the \$1,000,000 being there increases, and so he begins to believe that choosing both boxes is better. Let $S2$ denote the presence of the \$1,000,000. Formally, as $P(A2)$ approaches $0$ or $1$, the conditional probabilities $P(S2|A1)$ and $P(S2|A2)$ approach $1$ and $0$, respectively. The value of $P(A2)$ where the expected utility of $A2$ and the expected utility of $A1$ are equal is where deliberation stops, and this is Tomas' final probability of two-boxing.  On Skyrms' model this does not in fact end in a reconciliation of evidential and causal decision theory. Supposing Tomas is an evidentialist and begins on the fence, he ends deliberation most probably one-boxing, but also attaches some positive probability to two-boxing.

To this, Eells (\cite{eells1984}) introduces a model called ``continual conditional expected utility maximization'' which embraces Skyrms' idea that deliberation generates information upon which we should update our expected utilities but also introduces the notion that agents may face an urgency to act. Thus, depending on whether one wants to reach a decision quickly, one might eschew the states of indecision that Skyrms claims the evidentialist stuck in. Eells believes this reconciles the prescriptions of evidential and causal decision theory on Newcomb's problem, resulting in two-boxing. However, as Huttegger (\cite{huttegger2023}) rightly points out, this is a large deviation from traditional evidential decision theory. Whether an agent rushes to a decision or procrastinates are features of an agent not well captured by her preferences. Therefore, the proposed solution arguably fails.

Huttegger takes a different approach to reconciliation in light on Skyrms' findings. His amendment to Skyrms' model is a relaxation of the assumption that as $P(A2)$ approaches $0$ or $1$, the conditional probabilities $P(S2|A1)$ and $P(S2|A2)$ approach $1$ and $0$. That is, conditional probabilities of the states given acts are not functions of our choice probabilities. Indeed, in the original Eellsian account, there is nothing over and above one's informed beliefs and desires upon which the agent's decision is based; convergence towards one or the other act is not required for the appropriate screening off. Instead, conditional probabilities change by a separate ``independence dynamics'' as a function of time, or stages, in the deliberational process, moving closer to one another over the course of deliberation.\footnote{One may argue against deliberation generating such information for the agent. However, for the purpose of my current analysis, I leave aside these issues. See \cite{huttegger2023} for a discussion.} The independence dynamics is formally defined as follows.\footnote{If $P(A1) = 0$, then $P(S2| A1)$ is not well defined. Huttegger states this obstacle can be overcome by requiring that dynamics of $P(S2|A1)$ is continuous with the dynamics for arbitrarily close states that have $P(A1) > 0$.} 

$$
\frac{dP(S2|A2)}{dt} =
\begin{cases}
\text{positive if} \hspace{1mm} P(S2|A1) > P(S2|A2)\\
\text{negative if} \hspace{1mm} P(S2|A1) < P(S2|A2)\\
\end{cases}
$$

Likewise,

$$
\frac{dP(S2|A1)}{dt} =
\begin{cases}
\text{positive if} \hspace{1mm} P(S2|A1) < P(S2|A2)\\
\text{negative if} \hspace{1mm} P(S2|A1) > P(S2|A2)\\
\end{cases}
$$

There are also no reappearances of correlations, so

$$\frac{d[P(S2|A1) - P(S2|A2)]}{dt} = 0 \; \text{if} \; P(S2|A1) = P(S2|A2)$$

Under this dynamical process, evidential deliberation converges to two-boxing since the choice probability of two-boxing is governed by the adaptive dynamics when state and act are independent. It is precisely the introduction of the independence dynamics that brings us to this reconciliation. If the evidentialist does not believe her act is evidence for a state of the world, she in effect uses the same probabilities the causalist uses.

Furthermore, while in Skyrms' work, the end point of deliberation is where the choice probability of an act is 1 or $\frac{dP(A2)}{dt} = 0$, this is not the case under Huttegger's framework.\footnote{In Skyrms (\cite{skyrms1982}), that the adaptive dynamics continues until the probability of an act equals 1, and does not exceed 1, is guaranteed by the fact that this is when deliberation ends. This is not the case under Huttegger's framework; deliberation does not end when the probability of an act reaches 1. Therefore, as stated here, it is possible that $P(A2)$ exceeds 1 since the rule that the change in choice probability is proportional to the difference in expected utility does not ensure that $P(A2)$ remains within the probability simplex. As such, we stipulate that the adaptive dynamic rules which are permissible under this general formulation are those which effectively slow as they reach the boundary, therefore remaining within the probability simplex over the course of deliberation.} Rather, deliberation, in most cases, will continue until $\frac{dP(A2)}{dt} = 0$ \textit{and} the agent reaches state-act independence. I say ``in most cases'' since Huttegger does not assume deliberation always leads to full state-act independence. This is because deliberation can sometimes fail provide all the information we need, for example, if the agent believes that the predictor in Newcomb’s problem knows more about how he makes decisions than he knows about himself. If this is so, there are hidden factors influencing his choice which he cannot access via deliberation. Nonetheless, Huttegger states that situations where agents' acts are determined solely on the basis of their desires, beliefs and decision rule are the ``most natural setting for decision theory'' (\cite[22]{huttegger2023}). As such, I will be considering those cases in which the agent's deliberative process is sufficient to screen off state-act correlations. 

In Huttegger's framework, the reason that the independence dynamics can continue after the adaptive dynamics concludes is because the operation of the independence dynamics is independent of the adaptive dynamics: it is not a function of the agent's choice probabilities. It is important to note that, on this interpretation, the relative strength of the independence and adaptive dynamics becomes relevant to where the agent ends deliberation. Huttegger's work finds that the exact specification of the operation of the independence dynamics relative to the adaptive dynamics  does not matter for Eells' reconciliation project on Newcomb's problem. In this paper, I show that it does matter for other decision problems on which evidential and causal decision theory diverge.

I will not reconstruct Huttegger's work on Newcomb's problem here but rather apply his same framework to Death in Damascus. I begin by determining the dynamics on the boundaries and discuss the more complicated interior dynamics in Sections 5 and 6.

\section{Death in Damascus for the deliberative evidentialist}

In the language of metatickles, both Tereza's act of staying or fleeing and Death's appointment in Damascus or Aleppo are effects of a common cause; that is, the cognitive architecture of the agent upon which Death bases his appointment, sometimes referred to as the agent's ``type'' (\cite{joyce2018}). Thus, conditional on the metatickle, $T$, which fully captures Tereza's beliefs and desires, states and acts are independent, and knowledge of the beliefs and desires of the kind caused by the common cause screens off the evidence that her choice provided for Death's location. Without making reference to common causes, but noting that deliberation can screen off state-act correlations, Huttegger introduces the independence dynamics, which, along with the adaptive dynamics  describes the changes in an agent's choice probability over the course of her deliberation.

\begin{figure}[h]
    \centering
    \includegraphics[scale=0.45]{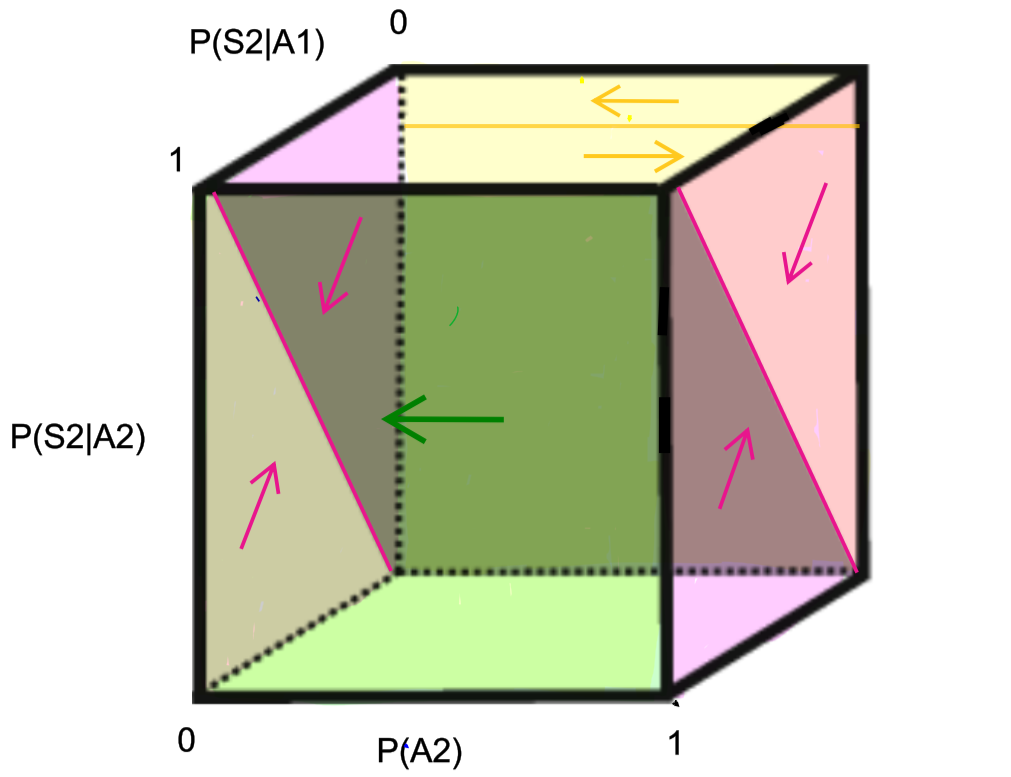}
    \caption{Deliberative evidentialist reasoning under Huttegger's framework}
    \label{fig:huttegger}
\end{figure}

Under Huttegger's framework, $P(S2|A2)$ and $P(S2|A1)$ may vary independently so the deliberational space is represented in three dimensions; one being $P(S2|A1)$; the other  $P(S2|A2)$; and the final being Tereza's probability of fleeing, $P(A2)$, all of which change during the deliberative process. The deliberational space is depicted in Figure~\ref{fig:huttegger}. Note that the cube does not represent a phase diagram as the magnitude of the movement in any particular direction has not been specified. It should rather be thought of as a qualitative tool by which we may analyse where deliberation leads us.

Recall the conditional expected utility formulae of evidential decision theory. That is,

$$EU_{evid}(A1) = D(S1 \& A1)P(S1|A1) + D(S2 \& A1)P(S2|A1)$$
$$EU_{evid}(A2) = D(S1 \& A2)P(S1|A2) + D(S2 \& A2)P(S2|A2)$$

Given these formulae and the logical fact that $P(S1|A1) + P(S2|A1) = 1$ and  $P(S1|A2) + P(S2|A2) = 1$ (one or other state of the world must obtain given our act), we may discern the movement of $P(A2)$ on the faces of the cube by calculating the expected utility of both acts. First, let us address the front face, indicated in green, where $P(S2|A1) = 1$. The top edge is where $P(S2|A2) = 1$. Here we have $EU(A1) = 10$ and $EU(A2) = -1$. Since $EU(A2) < EU(A1)$, by the adaptive dynamics, $P(A2)$ decreases. Similarly, on the bottom edge of the front face, where  $P(S2|A2) = 0$, $EU(A2) < EU(A1)$.

It can be verified that all points in between the edges also lead to a final choice probability of $P(A2) = 0$ on the front face of the cube. This is intuitive as, if $P(S2|A1) = 1$, Tereza can outsmart Death. That is, if the probability of Death being in Aleppo given that Tereza stays in Damascus is 1, she should surely stay in Damascus and not pay the extra 1 util to flee.

Now consider the back face, indicated in yellow, where $P(S2|A1) = 0$. The top edge is where $P(S2|A2) = 1$. Here we have $EU(A1) = 0$ and $EU(A2) = -1$. Again $P(A2)$ decreases. However, on  the bottom edge of the back face, the dynamics look different. Here,  $P(S2|A2) = 0$, so $EU(A1) = 0$ and $EU(A2) = 9$. Since $EU(A2) > EU(A1)$, $P(A2)$ increases. The exact point at which Tereza prefers fleeing over staying will be explored in the next section using what I call the \textit{plane of indifference}.

However, we have not yet considered the operation of the independence dynamics on the left and right faces, indicated in pink. This leads us to what Huttegger calls the Eells-Jeffrey manifold, represented by the grey diagonal face in the cube, and consists of all points where $P(S2|A2) = P(S2) = P(S2|A1)$, in other words, where there is state-act independence. Movement toward the Eells-Jeffrey manifold is given by the evolving metatickle which screens off states from acts during an agent's deliberation. If our metatickle is sufficient to reach full state-act independence, we must determine the movement on the manifold itself.

\begin{figure}[h]
    \centering
    \includegraphics[scale=0.5]{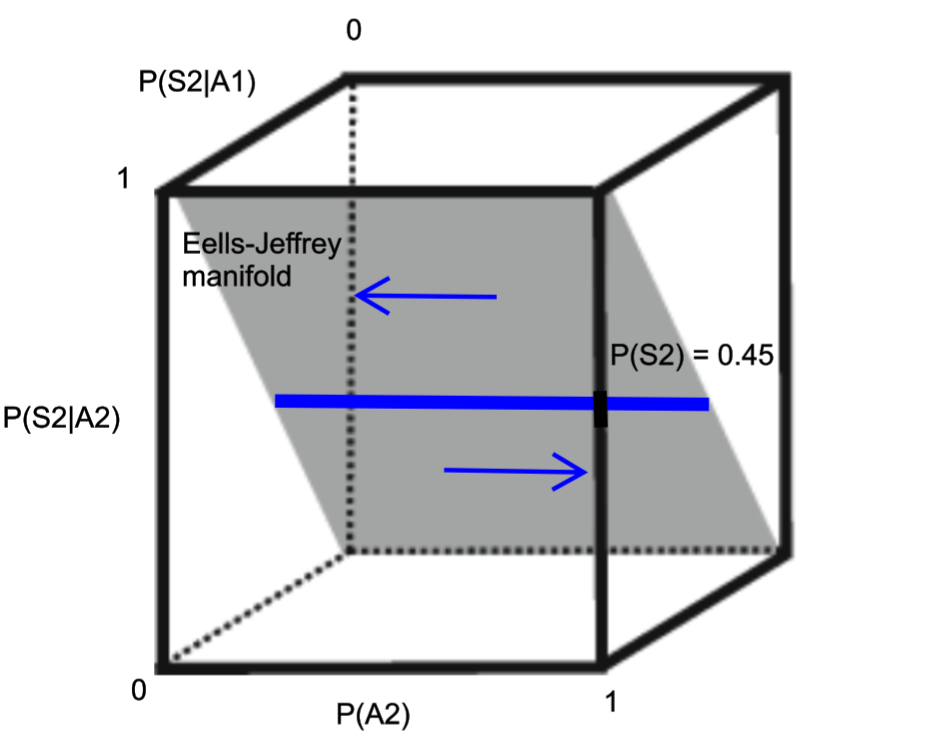}
    \caption{Evidentialist reasoning on the Eells-Jeffrey manifold}
    \label{fig:eellsjeffrey}
\end{figure}

All areas above the bold blue line move to $P(A2) = 0$ and all areas below it move to $P(A2) = 1$ by the adaptive dynamics. The bold blue line is where $P(S2|A2) = P(S2|A1) = 0.45$. Here, $EU(A1) = EU(A2) = 4.5$ so there is no movement in $P(A2)$ as per our specification of the adaptive dynamics. I have not yet discussed the dynamical movement in much of the interior of the cube, which is the subject of the next section, but first it is worth noting the following facts.

Here, we have multiple equilibria represented by the bold blue line. All of these choice probabilities of $P(A2)$ render the expected utility of staying equal to that of fleeing, despite the fact that the unconditional probability of Death being in Damascus is $0.45$.\footnote{It should be noted that such lines of equilibria in general exhibit structural instability. That is, they are sensitive to changes in the dynamical rule (\cite{skyrms1990}} However, this is also the case for the deliberative causalist. Though the mixed act of fleeing with probability $0.474$ is the end point of deliberation, at this point, all other acts have equal expected utility so all are equally permissible (\cite{joyce2012}). Here, one might inquire what then renders the mixed act the correct answer. The reason is that this is the uniquely ratifiable act (should one have the option to execute it using a chance device that represents this probability distribution). That is, it is the only act where, upon knowledge that one has chosen it, one would not prefer otherwise.\footnote{This is also supported by consideration of the fact that the mixed act would constitute the Nash equilibrium of a normal form game with Death and Tereza as players. For discussion of the connection between ratifiability in deliberative decision theory and Nash equilibria in game theory, see \cite{skyrms1990}; \cite{harper1986}; \cite{joycegibbar1998}; and \cite{weirich2016}.} 
 
In Sections 5 and 6, I show that the prescription of the mixed act under Huttegger's framework hinges upon two further conditions: (i) the independence dynamics does not take the ``shortest path'' to state-act independence, and (ii) the relative strength of the adaptive and independence dynamics must be such that they reach the Eells-Jeffrey manifold exactly where $P(A2) =  0.474$. Since these conditions imply that deliberation must proceed via a very specific route to the precise choice probability, it will not deliver reconciliation under many plausible specifications of the deliberative process. First, I consider what happens under one plausible specification of the independence dynamics.

\section{Shortest-path independence and the plane of indifference}

In this section, I offer an original analysis of the Huttegger's deliberative framework given a plausible version of the dynamical process, which I call the \textit{shortest-path independence dynamics}. I prove the existence of what I call the \textit{plane of indifference} which determines why the framework is irresolute in the case of Death in Damascus and not in Newcomb's problem. I then show that, under Huttegger's framework, this plane of indifference exists in all two-act, two-state decision problems which exhibit the basic mathematical structure of either Newcomb or decision instability problems. The upshot is that the precise specification of the independence dynamics matters for reconciliation only depending on the positioning of this plane of indifference. This provides a principled way of knowing \textit{ex-ante} whether a reconciliation of evidential and causal decision theory is plausible for a wide range of decision problems under this framework.

Informally, the independence dynamics drives the agent's conditional probabilities toward one another over time, though the exact way in which this occurs is left open in Huttegger's work. One way the independence dynamics could operate is by adjusting one starting conditional probability to match the other. For example, if Tereza's initial value of $P(S2|A2)$ is $0.99$ and her initial value of $P(S2|A1)$ is $0.01$, she adjusts up the value of $P(S2|A1)$ until it also equals $0.99$. However, this does not seem particularly rational. Given the description of the decision problem, both of her initial conditional probabilities reflect Death's reliability in predicting her action, so there appears no reason to count one rather than the other as more viable for informing her unconditional credence in the state of the world.

\begin{figure}[h]
    \centering
    \includegraphics[scale=0.4]{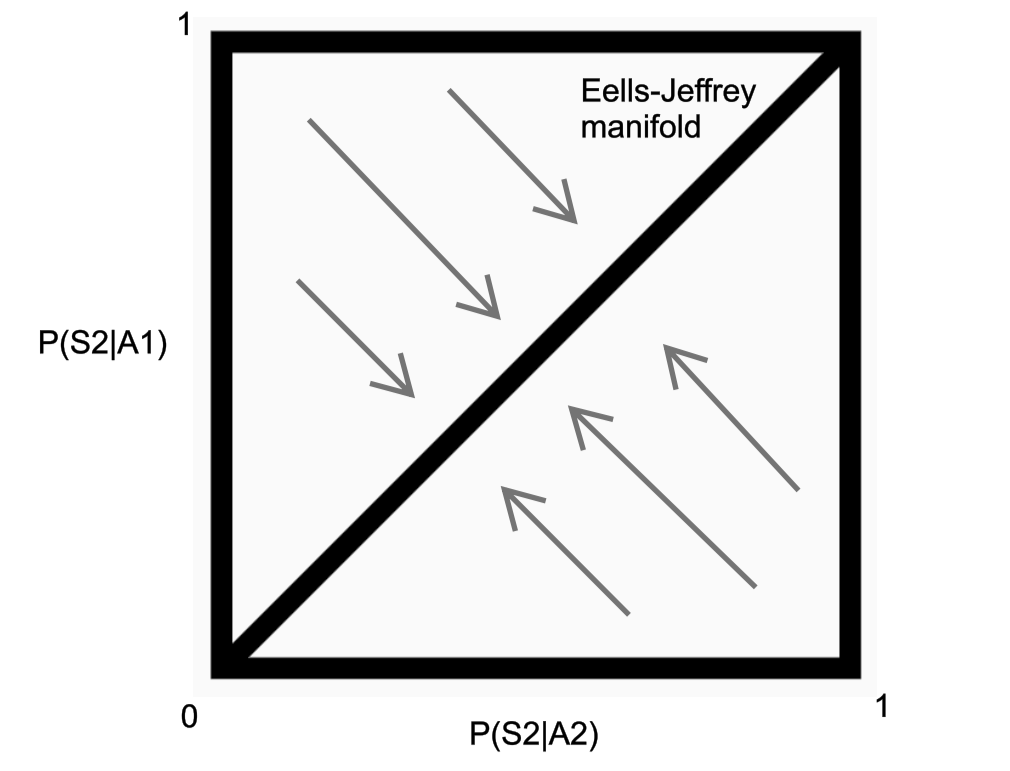}
    \caption{Shortest path to Eells-Jeffrey manifold}
    \label{fig:shortpath}
\end{figure}

A more plausible version of the independence dynamics would be one that concludes at the average across her two initial conditional probabilities. Since a movement in the direction of the manifold for one conditional probability then implies an equal movement in the direction of the manifold for the other, the independence dynamics decrees -- absent its interaction with the adaptive dynamics --  that Tereza's conditional probabilities move in the straight line that captures the shortest path to the manifold. This is illustrated in Figure~\ref{fig:shortpath}, which represents a slice through the dynamical cube and the diagonal line represents the manifold.\footnote{Since the analysis is qualitative, this may extend to sufficiently similar independence dynamics, though this has not yet been considered.}

To see what this means for our deliberative process, first we must return to an important feature of the dynamical cube previously overlooked. In our earlier illustration, the line of equilibria on the manifold represented a situation where there was no movement prescribed by the adaptive dynamics; any choice probability of $P(A2)$ was acceptable since all mixtures of acts had equal expected utility. Moving off the Eells-Jeffrey manifold, we see that this is not only a feature existing at state-act independence but, as I will show, there exists a whole plane on which the adaptive dynamics prescribes no change in $P(A2)$. This occurs where the two conditional probabilities of state given act, $P(S2|A1)$ and $P(S2|A2)$ sum to $0.9$. The fact that this is a plane of the cube follows from the fact that two axes of the 3-dimensional space represent these conditional probabilities. The fact that the  adaptive dynamics decrees no change in choice probability on this plane can be seen from the following.

Let $P(S2|A1) + P(S2|A2) = 0.9$ and note it is true by definition that $P(S1|A1) = 1 - P(S2|A1)$ and $P(S1|A2) = 1 - P(S2|A2)$. Then

\begin{align*}
EU(A1) &= 0P(S1|A1) + 10P(S2|A1) \\
&= 10P(S2|A1)
\end{align*}
And 

\begin{align*}
EU(A2) &= 9P(S1|A2) -1P(S2|A2) \\
&= 10P(S2|A1)
\end{align*}

\begin{figure}[h]
    \centering
    \includegraphics[scale=0.55]{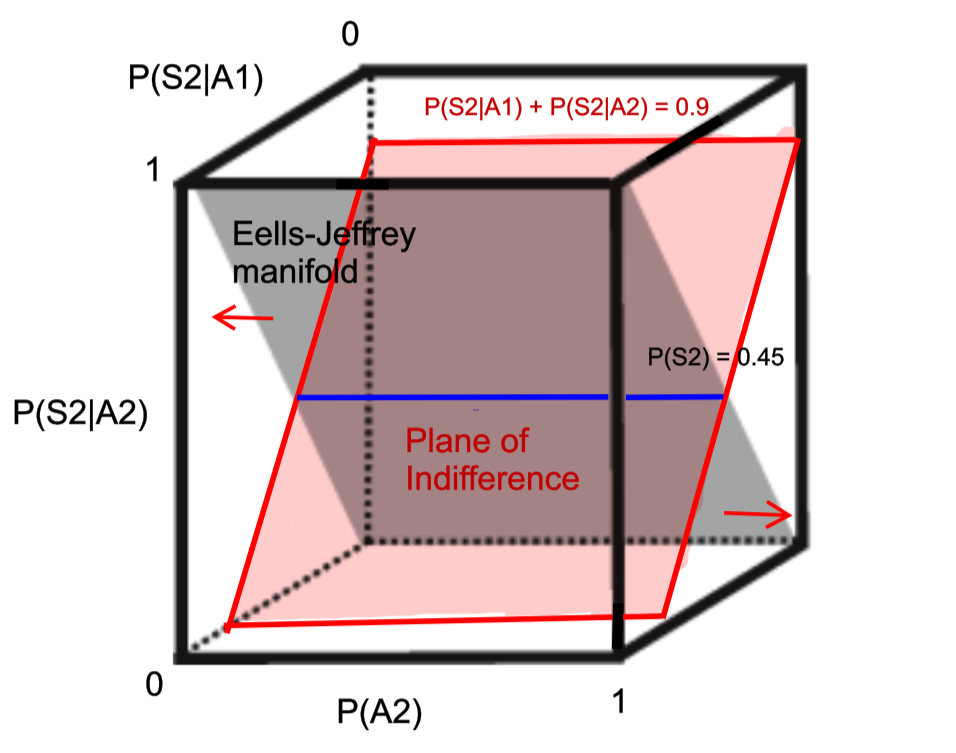}
    \caption{The plane of indifference}
    \label{fig:indifference}
\end{figure}

Since the expected utility of both acts are equal as defined in terms of $P(S2|A1)$, the adaptive dynamics prescribes no movement on the plane given by $P(S2|A1) + P(S2|A2) = 0.9$. Figure~\ref{fig:indifference} illustrates what I call the \textit{plane of indifference}.

The key feature of this plane is that if one begins deliberation on the plane, since $P(A2)$ does not change, one simply moves by the independence dynamics toward the line of equilibria and ends deliberation with the same choice probability as she began with. Of utmost interest is what happens when we begin deliberation either below or above the plane of indifference. It turns out that if Tereza begins at any point below the plane, where $P(S2|A1) + P(S2|A2) < 0.9$, Tereza's deliberation concludes that she should flee to Aleppo with probability $1$. If she begins above the plane, where $P(S2|A1) + P(S2|A2) > 0.9$, Tereza concludes she must stay in Damascus, and flee to Aleppo with probability $0$.

For example, consider $P(S2|A1) + P(S2|A2) = 1$. Here, we have a 2-dimensional plane which sits above the plane of indifference. All initial choice probabilities will lead Tereza to staying. To see this, note that since we have imposed the constraint $P(S2|A2) + P(S2|A1) = 1$, and by logical fact, $P(S1|A1) + P(S2|A1) = 1$ and  $P(S1|A2) + P(S2|A2) = 1$, our constraint implies $P(S1|A2) + P(S1|A1) = 1$. Given these formulae, we may calculate our expected utilities. First, consider the top edge of the plane, where $P(S2|A2) = 1$. We see that $EU(A1) = 0$ and $EU(A2) = -1$. Since $EU(A2) < EU(A1)$, by the adaptive dynamics, $P(A2)$ must reduce. Similarly, on the bottom edge of the plane where  $P(S2|A2) = 0$, $EU(A2) < EU(A1)$. Since $P(S2|A2) + P(S2|A1) = 1$, shortest-path independence dynamics drives her unconditional probability $P(S2)$ to $0.5$. In the middle of the plane on its intersection with the Eells-Jeffrey manifold, $EU(A1) = 5$ and $EU(A2) = 4$ so, again, $EU(A2) < EU(A1)$. As a result, deliberation moves Tereza toward staying in Damascus until we reach a stable equilibrium point where $P(A2) = 0$ and $P(S2|A2) = P(S2) = P(S2|A1) = 0.5$. Analogous reasoning applies when we begin on the other side of the plane and $P(S2|A1) + P(S2|A2) < 0.9$.

In what follows, I will prove that the adaptive dynamics is governed by whether we are below or above the plane of indifference for a general payoff table representing a wide range of decision instability problems. Let $a$ denote the utility assigned to survival and $b$ the utility assigned to death. Since we consider an asymmetric payoff table, let $c$ denote the cost of fleeing. Our payoff table represents a general version of a wide range of asymmetric decision instability problems where $a > b$ and $c \leq a - b$. Other problems with a similar structure are the Murder Lesion problem and the Psychopath Button (\cite{egan2007}; \cite{arntzenius2008}; \cite{joyce2012}).

\begin{table}[h]
    \centering
    \begin{tabular}{ l | l | l}
    & S1 & S2 \\
    \hline
    A1 & b & a \\
    \hline
    A2 & a - c & b - c \\
\end{tabular}
    \caption{Generalised payoff table for asymmetric decision instability problem}
\end{table}

The plane of indifference can be defined in terms of the utilities in the payoff table. Recall that the adaptive dynamics prescribes no movement in $P(A2)$ when $EU(A1) = EU(A2)$. This is when
$$bP(S1|A1) + aP(S2|A1) = (a-c)P(S1|A2) + (b-c)P(S2|A2)$$ 

By substitution and rearranging, we get 

$$ P(S2|A1) + P(S2|A2) = \frac{a-b-c}{a-b}$$

We must prove that the sum is defined and that it is greater than or equal to 0 and less than or equal to 2 in order for it to appropriately represent an agent's conditional probabilities. First, by definition of the payoff table $a > b$, so the denominator is positive and the expression is defined. Second, $\frac{a-b-c}{a-b} \geq 0$ entails that the numerator is also positive. Note that since $a > b$, this will be satisfied as long as $c \leq a-b$. Of course, this is true from the definition of the asymmetric decision instability problem. If the cost of fleeing was greater than the difference between survival and death, we would not be in a case of asymmetric Death in Damascus as it would never be preferable to flee. Finally, $\frac{a-b-c}{a-b} \leq 2 = a-b-c \leq 2(a-b) = -c \leq a - b$. This is satisfied by definition of the payoff table again, as $c$ is positive and $a > b$ so the left hand side is negative whilst the right is positive.

From this equation for the plane of indifference, we can see that as the cost of fleeing increases, the right hand side of the equation reduces, meaning the plane of indifference will move downwards in the diagonal space of the dynamical cube. This decreases the area of the cube where Tereza's deliberation leads her to flee. In other words, the greater the cost of fleeing, the more sure Tereza must be that Death is in Damascus than that he is in Aleppo in order that rationality decree she purchases the ticket to flee.\footnote{If $c = 0$, we are in a symmetric decision instability problem where the plane of indifference intersects the Eells-Jeffrey manifold at $P(S2) = 0.5$.} Now that we have proved the existence of an indifference plane, we can demonstrate how the adaptive dynamics will operate either side of it in a general setting.

Since $a-b$ is positive (the utility of living exceeds that of dying) we can easily replace our equalities in the above existence proof with inequalities. The direction of the inequality does not change throughout the proof. It follows that:

$$P(S2|A1) + P(S2|A2) > \frac{a-b-c}{a-b} \iff EU(A1) > EU(A2)$$
$$P(S2|A1) + P(S2|A2) < \frac{a-b-c}{a-b} \iff EU(A1) < EU(A2)$$

This means that if the agent begins deliberation above the plane, she will end deliberation with $P(A2) = 0$ and if she begins below it, she will end deliberation with $P(A2) = 1$.

Here, one might ask whether her dynamical deliberation could cross over the plane. In principle, it could. However, this would be to violate the plausible stipulation we have made that the ideal deliberator approaches the Eells-Jeffrey manifold via the shortest-path independence dynamics. By definition of how I have specified the shortest-path dynamics, the path toward the manifold is perpendicular to the manifold. This can be seen in Figure~\ref{fig:shortpath}. We can also prove that the indifference plane is perpendicular to the manifold by showing that the dot product of the normal vectors of both planes is 0. Since the normal vector of a plane is perpendicular to it, it is sufficient to show that the normal vectors are perpendicular to each other in order to show that the planes are perpendicular. The plane of indifference is given by $P(S2|A2) + P(S2|A1) =  \frac{a-b-c}{a-b}$ and the Eells-Jeffrey manifold is given by $P(S2|A2) - P(S2|A1) = 0$. The normal vectors are therefore $A = \langle 1, 1 \rangle$ and $B = \langle 1, -1 \rangle$. The dot product is thus $A \cdot B = 0$. The planes are therefore perpendicular and this will hold for any value of $\frac{a-b-c}{a-b}$.

It is clear, therefore, that the shortest-path dynamics decrees dynamical adjustments of conditional probabilities that run parallel to the plane of indifference and do not cross it. Given this feature, one's initial starting point entirely determines the ending point of deliberation. This is true of more general cases than the one considered here, as long as the payoff table bears the same mathematical relationship to the one presented above, where $a > b$ and $c \leq a - b$, and raises important questions for the reconciliation of causal and evidential decision theory for problems of decision instability in Huttegger's deliberative framework.

Now let us consider why this problem does not arise in Newcomb's problem. In short, the reason is that the structure of the payoff table renders the plane of indifference \textit{parallel} to the Eells-Jeffrey manifold. This means that, above or below the plane, shortest-path independence dynamics will necessarily pass through it to the Eells-Jeffrey manifold where adaptive dynamics dictates that Tomas takes both boxes. Consider the following generalised payoff table where $a > b$ and $c \leq a-b$. Other problems with a similar structure are the Cholesterol problem, Smoking problem, and Solomon's problem (\cite{skyrms1980}; \cite{gibbardharper1978}; \cite{eells1982}).

\begin{table}[h]
    \centering
    \begin{tabular}{ l | l | l}
    & S1 & S2 \\
    \hline
    A1 & b & a \\
    \hline
    A2 & b + c & a + c \\
\end{tabular}
    \caption{Generalised payoff table for Newcomb's Problem}
\end{table}

As above, the plane of indifference is found where $EU(A1) = EU(A2)$. This is when

$$bP(S1|A1) + aP(S2|A1) = (b+c)P(S1|A2) + (a+c)P(S2|A2)$$

By substituion and rearranging, we get 

$$P(S2|A1) - P(S2|A2) = \frac{c}{a-b}$$

We must prove that the difference is defined and that it lies between -1 and 1 inclusive in order for it to appropriately represent an agent's conditional probabilities. First, by definition of the payoff table $a > b$, so the denominator is positive and the expression is defined. Second, $-1 \leq \frac{c}{a-b} = b - a \leq c$. This is satisfied by definition of the Newcomb payoff  table, since if $c$ was strictly less than $b-a$, $c$ would be negative, and there would be no benefit to two-boxing. Finally, $\frac{c}{a-b} \leq 1 = c \leq a -b$. This is again satisfied by the definition of Newcomb payoffs, since if $c$ were strictly greater than $a-b$, this would mean $c + b > a$ and it would therefore always be better to two-box.

\begin{figure}[h]
    \centering
    \includegraphics[scale=0.5]{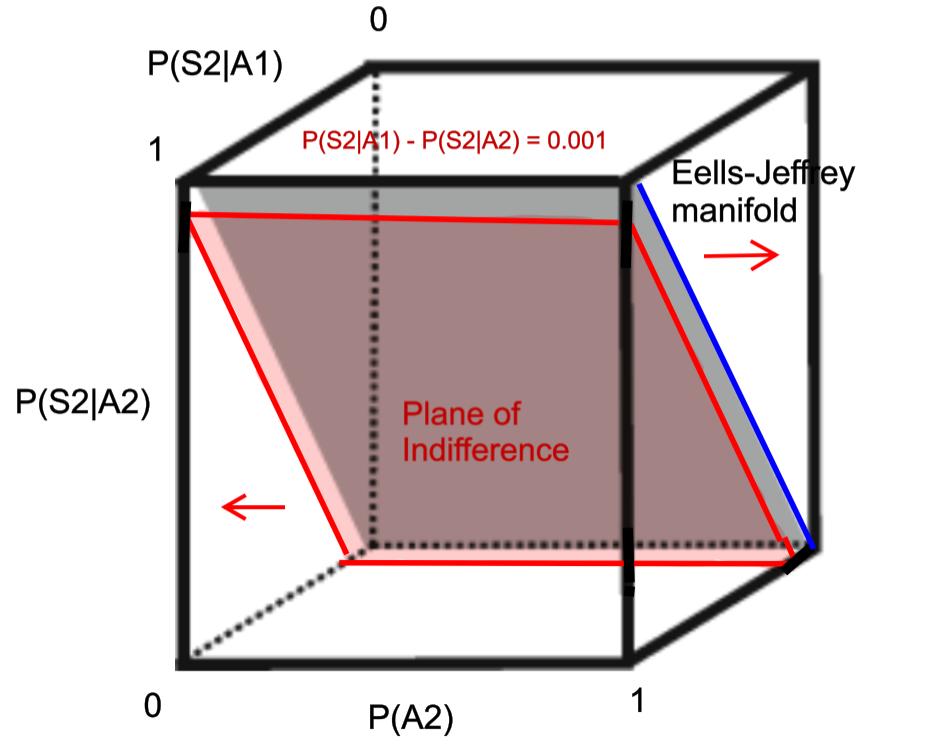}
    \caption{The plane of indifference for the Newcomb problem}
    \label{fig:plane-of-indifference2}
\end{figure}

Notice here that the relationship that defines the plane is not a sum but a difference. This means that the plane is parallel to the Eells-Jeffrey manifold. This is easily proved by taking the ratio of the components of their normal vectors and showing that they are the same. Indeed, they are both 1. This will hold for any value of $\frac{c}{a-b}$. It will be illuminating to rewrite the above condition as $P(S2|A2) = P(S2|A1) - \frac{c}{a-b}$ so we see the indifference plane sits below the manifold. This is illustrated in Figure~\ref{fig:plane-of-indifference2}.

The movement decreed by the adaptive dynamics on either side of the plane in the Newcomb problem is given by examining the following biconditional statements. As before, the proof proceeds straightforwardly from the existence proof replacing the equalities with inequalities without any change in direction, as the term $a-b$ is positive.

$$P(S2|A1) > P(S2|A2) + \frac{c}{a-b} \iff EU(A1) > EU(A2)$$
$$P(S2|A1) < P(S2|A2) + \frac{c}{a-b} \iff EU(A1) < EU(A2)$$

We can see from Figure~\ref{fig:plane-of-indifference2} that when $P(S2|A1) > P(S2|A2)$, we are below the Eells-Jeffrey manifold. So when $P(S2|A1) > P(S2|A2) + \frac{c}{a-b}$, we are below the plane of indifference. Here, the biconditional statements above reveal that the rational act according to our adaptive dynamics is to one-box. By analogous reasoning, all points above the indifference plane end deliberation in two-boxing. As the independence dynamics moves the agent towards the Eells-Jeffrey manifold, and the Eells-Jeffrey manifold lies above the indifference plane, the adaptive dynamics decrees that the agent ought to two-box in Newcomb's problem, corroborating Huttegger's conclusion. 

As the value of $c$, the monetary sum under the transparent box, increases, the plane of indifference shifts downward in diagonal space away from the Eells-Jeffrey manifold. As a result, the region of the cube where Tomas should rationally one-box reduces. This is intuitive as, by description of the problem, the agent only receives the value $c$ when he two-boxes, so the greater the value of $c$, the greater the incentive to two-box. The denominator $a-b$ captures the difference between the contents of the opaque box in the two states of the world. If this difference is large, the plane shifts upwards, expanding the region of points which decree as rational one-boxing. This again is intuitive, as the greater the incentive to one-box, the less sure the agent need be that the predictor put $a$ there in order for him to rationally choose it. Note that when $c = 0$ the plane of indifference is exactly equivalent to the Eells-Jeffrey manifold. It might be tempting to think that if there is nothing under the transparent box, the agent should one-box, but this is not the correct answer. Recall that when we have reached state-act independence, Tomas does not see his act as evidence about the state of the world, so he is rationally indifferent between one-boxing and two-boxing. The causalist answer is the same, as the payoffs are the same under both states of the world. 

The preceding discussion has shown that it is the plane of indifference which determines rational action in both decision problems. The crucial difference, however, is that regardless of the exact specification of the independence dynamics, the agent's trajectory of deliberation in Newcomb's problem may pass through the indifference plane to the Eells-Jeffrey manifold, since the two are parallel. This means that where one begins deliberation does not determine where one ends in the same way that it does in the Death in Damascus problem. Here, if we accept the plausibility of the shortest-path independence dynamics, movement toward the manifold never crosses the indifference plane, since the independence path and plane of indifference are parallel to one another. This analysis shows that the relatively straightforward reconciliation of causal and evidential deliberation for Newcomb's problem under Huttegger's deliberative framework is not so straightforwardly achieved in problems of decision instability. Much more would have to be said on the nature of the independence dynamics in order to determine whether we may cross the plane of indifference and end deliberation with a resolute answer. In the next section, I turn to these further requirements.

\section{On the possibility of reconciliation}

Recall that, under Huttegger's framework, deliberation ends when the adaptive dynamics prescribes no further movement and when we reach state-act independence. In this section, I show that this will only lead to a reconciliation under two very specific conditions: (i) the independence dynamics must be specified such that it does not take the shortest path to the manifold, and (ii) the adaptive dynamics and independence dynamics must have a relative speed such that they reach the Eells-Jeffrey manifold at precisely the point of reconciliation.

As we saw from the previous section, if we take the shortest-path independence dynamics to be true, whether Tereza begins above or below the plane of indifference determines where she will end deliberation. The only time, therefore, where she could end deliberation with $P(A2) = 0.474$ is when she begins with deliberation with her choice probability at $P(A2) = 0.474$ and her conditional probabilities precisely on the plane of indifference (where they sum to $0.9$). In this case, shortest-path independence will move her directly to the line of equilibria without any change in her choice probability. This is a case where there appears to be no deliberation at all driving her conclusion, and is therefore implausible as a reconciliation of evidential and causal decision theory via deliberation.

\begin{figure}[h]
    \centering
    \includegraphics[scale=0.55]{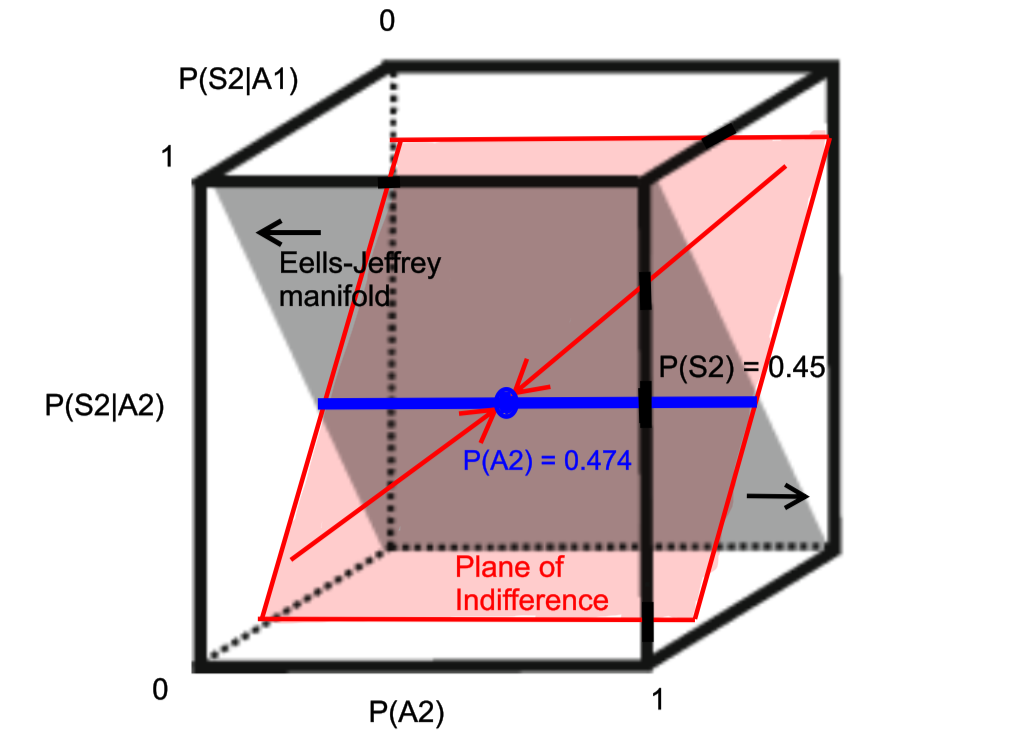}
    \caption{Diagrammatic portrayal of the deliberative evidentialist reasoning under Huttegger's framework. The red arrows represent possible trajectories to reconciliation. Both trajectories cross the plane of indifference where the upper red arrow begins above it and the lower red arrow begins below it.}
    \label{fig:evidentialist-huttegger}
\end{figure}

Of course, there may be viable independence dynamics other than shortest-path independence so let us relax this assumption. However, even if we allow violation of shortest-path independence, it must be the case that the relative speed of the adaptive and independence dynamics is such that the agent reaches the Eells-Jeffrey manifold precisely at the point where it intersects the plane of indifference at $P(A2) = 0.474$. If Tereza reaches the manifold on the equilibrium line at any point to the left or right of this, $P(A2) \neq 0.474$ and $\frac{dP(A2)}{dt} = 0$ so we do not achieve reconciliation. If Tereza reaches the manifold at any other point above or below the equilibrium line, the adaptive dynamics leads her to $P(A2) = 0$ or $1$ depending on whether this is above or below the plane of indifference. It is only if the two conditions I have specified obtain that we may witness trajectories such as those depicted in Figure~\ref{fig:evidentialist-huttegger}, but the reconciliation here appears forced.

Again, it is important to recognise that this was not a issue in the case of Newcomb's problem. Here, regardless of the specification of the independence dynamics, since the Eells-Jeffrey manifold lies on the side of the indifference plane where two-boxing is rational, as long as deliberation leads us to state-act independence, the framework will always prescribe the correct answer. The relative strength of the independence and adaptive dynamics may lead Tomas to different points on the line of equilibria where the Eells-Jeffrey manifold intersects the right face of the cube, but this does not change Tomas' ultimate action, as $P(A2) = 1$. Where he concludes deliberation only determines his beliefs about his winnings. That is, he believes himself to be more fortunate if he ends deliberation where the probability of the \$1,000,000 being there, $P(S2)$, is high, and less fortunate if he ends deliberation where it is low.

The analysis I have offered in this section therefore represents a principled way to delineate when the specification of the independence dynamics matters for the reconciliation of evidential and causal decision theory under Huttegger's framework. In particular, it depends on whether the plane of indifference intersects the Eells-Jeffrey manifold or not. If it does not, implying it lies entirely to one side of the Eells-Jeffrey manifold, the specification of the independence dynamics does not matter. Any independence dynamics that moves the agent in the direction of state-act independence over time will lead to the same answer. As is shown from the generalised proofs, for any problem representing the mathematical structure of the generalised Newcomb's problem, the plane of indifference will not intersect the Eells-Jeffrey manifold. For any problem representing the mathematical structure of the generalised Death in Damascus problem, the plane of indifference will be perpendicular to the Eells-Jeffrey manifold, and the specification of the independence dynamics as well as its strength relative to that of the adaptive dynamics, matters for where the agent concludes deliberation. We therefore have a robust way of determining \textit{ex-ante} whether reconciliation of evidential and causal decision theory is plausible for a wide range of two-state, two-act decision problems under this framework.

Note that what is important is not whether the plane of indifference is perpendicular or parallel to the Eells-Jeffrey manifold, but whether it \textit{intersects} the manifold, meaning that the analysis here could in principle be extended to other decision problems, where the angle of the plane of indifference relative to the manifold differs, in order to determine whether specification of the independence dynamics matters in these problems. Furthermore, we would expect the key result -- that the relative strength of the adaptive and independence dynamics matters for reconciliation -- to hold in larger ($n x n$) decision problems, though this has not as yet been investigated.

\section{Conclusion}

The prescriptions of evidential and causal decision theory come apart in two general classes of problems known as Newcomb problems and decision instability problems. Huttegger (\cite{huttegger2023}) has developed a framework for evidential deliberation building on Eells' (\cite{eells1982}) metatickle approach and Skyrms' (\cite{skyrms1984}) deliberation dynamics which reconciles the prescriptions of the evidentialist and causalist in Newcomb's problem. Since deliberation results in increasing awareness of our beliefs and desires (and these are the mechanisms by which our action is determined), our acts no longer provide information about the state of the world. That is, deliberation screens off the state-act correlation which previously caused the evidentialist to choose the dominated act in Newcomb's problem. Huttegger's more sophisticated, deliberative evidentialist agent agrees with the causalist in preferring two-boxing.

In this paper, I have extended Huttegger's framework to consider an asymmetric case of decision instability: the Death in Damascus problem. I have shown that, in this context, Skyrms' adaptive dynamics and Huttegger's independence dynamics are insufficient to recommend a decisive answer. In Section 5, I consider a plausible version of the independence dynamics, shortest-path independence, and explore the particular features of the deliberative process that this independence dynamics decrees in Death in Damascus. We find that the dynamics decrees different answers for different initial starting points of deliberation. I prove the statements made here are applicable to a more general class of problems of decision instability, as long as the payoff table accords with some simple mathematical relationships. In particular, I show that there exists what I call a \textit{plane of indifference} where either act is equally acceptable, and this plane of indifference entails that where one concludes deliberation depends entirely on where one begins deliberation. This, however, is not true of the Newcomb case.

There are three upshots to this work. First, whilst application of the Eellsian metatickle to deliberation could straightforwardly lead to the correct answer in Newcomb's problem, this notion is not so easily extended to problems of decision instability, and the reconciliation requires assumptions that appear forced. Second, the proof of the plane of indifference for all two-state, two-act problems whose payoff tables exhibit the basic mathematical relationships in Section 5 provides us with a principled way of delineating those cases where the specification of the independence dynamics matters for a reconciliation of evidential and causal decision theory within this framework. Specifically, if the plane of indifference never intersects the Eells-Jeffrey manifold, the specification of the independence dynamics does not matter for reconciliation. If it does, reconciliation requires additional, and potentially questionable, assumptions about the exact specification of the adaptive and independence dynamics.

Finally, this work shows that the metatickle approach has so far failed to reconcile evidential and causal decision theory. Eells' and Jeffrey's original ideas were widely criticised for not providing details of how an agent arrives at knowledge of their own beliefs and desires, involving implicit assumptions, or idealisations that limit the metatickle approach (\cite{huttegger2023}; \cite{ahmed2014}; \cite{horwich1985}; \cite{joyce1999}; \cite{lewis1981a}; \cite{skyrms1982}; \cite{skyrms1984}). Attempts to resolve this using the theory of deliberation have shown it does not result in a reconciliation, rather the evidentialist is left in a state of indecision in Newcomb's problem (\cite{skyrms1990}). Eells' amendment (\cite{eells1984}) to his original idea then introduced spurious assumptions about other features of the agent, such as her felt urgency to act, which are marked deviations from traditional evidential decision theory (\cite{huttegger2023}). In this paper, I have shown that the most recent attempt to salvage Eells' idea, owing to Huttegger (\cite{huttegger2023}) also fails to deliver a reconciliation of evidential and causal decision theory in problems of decision instability. Future work on reconciliation would need to pay heed to the fact that our results will depend heavily on the interaction of the adaptive and independence dynamics, and any attempt at reconciliation would need to specify their relative strength such that evidential decision theory agrees with causal decision theory in both Newcomb and decision instability problems. 

\bibliographystyle{eptcs}
\bibliography{references}

\end{document}